\documentclass[10pt,a4paper]{article}
\usepackage[usenames]{color}
\usepackage{amsfonts}
\usepackage{amssymb}
\usepackage{amsmath}
\usepackage{latexsym}

\usepackage{epsfig}
\usepackage{graphicx}
\usepackage{bm}

\title{\bf \Large  Regular black holes in an asymptotically de Sitter universe }

\author{Jerzy Matyjasek$^1$\footnote{email: jurek@kft.umcs.lublin.pl}, 
Dariusz Tryniecki$^2$ and Mariusz Klimek$^1$
\\
\noalign{\vspace{3ex}}
\it{ \small $^1$Institute of Physics,  Maria Curie-Sk\l odowska University,}\\
\it{\small pl. Marii Curie-Sk\l odowskiej 1, 20-031 Lublin, Poland}
\\
\noalign{\vspace{3ex}}
\it{ \small $^2$Institute of Theoretical Physics,  Wroc\l aw University,}\\
\it{\small pl. Maxa Borna 9 , 50-204 Wroc\l aw, Poland}}

\date{}

\begin{document}

\maketitle
\begin{abstract}
A regular solution of the  system of coupled equations of the
nonlinear electrodynamics and gravity describing static and
spherically-symmetric black holes in an asymptotically de Sitter
universe is constructed and analyzed. Special emphasis is put on the
degenerate configurations (when at least two horizons coincide) and
their near horizon geometry. It is explicitly demonstrated that
approximating the metric potentials in the region between the horizons
by  simple functions and making use of a limiting procedure one obtains
the solutions constructed from maximally symmetric subspaces with
different absolute values of radii. Topologically they are
$AdS_{2}\times S^{2}$ for the cold black hole, $dS_{2}\times S^{2}$
when the event and cosmological horizon coincide, and the Pleba\'nski-
Hacyan solution for the ultraextremal black hole. A physically
interesting solution describing the lukewarm black holes is briefly
analyzed.
\end{abstract}

\vspace{.5cm}
\vspace{1cm}

\section{Introduction}	
It is a well known fact that the classical general relativity cannot
be trusted when curvature of the manifold approaches the Planck
regime. It means that understanding the nature of any classical
singularity that resides in the black hole interior as well as its
closest vicinity requires profound changes of the standard theory.
It is therefore natural that a great deal of effort has been concentrated on
construction of the singularity-free models.
(See for example Refs.~\cite{Sakharov,Gliner,Bardeen:68,frolov1,frolov2,Irina1,Borde,Mars,
abg,kiryll1,Dymnikova} and the references cited therein).

On of the most intriguing solutions of this type has been constructed
by Ayon-Beato and Garcia~\cite{abg} and by Bronnikov~\cite{kiryll1}
(ABGB). This solution to the system of coupled equations of the
nonlinear electrodynamics and gravity describes a regular static and
spherically-symmetric black hole characterized by the total mass
$\mathcal{M}$ and the magnetic charge $Q.$ The status of the nonlinear
electrodynamics in this model is to provide a matter source. The
casual structure of the solution is governed by the null geodesics
(``ordinary photons") rather than the photons of the nonlinear theory.
The latter move along geodesics of the effective
geometry~\cite{Salim1,Salim2}. 

The recent popularity of the models constructed within the framework
of the nonlinear electrodynamics has been stimulated by the fact that the latter 
appears naturally in the low-energy limit of certain
formulations of the string and M-theory~\cite{nat,cajtlin}.

The Lagrangian of the nonlinear electrodynamics adopted by Ayon-Beato and 
Garcia and by Bronnikov has Maxwell asymptotic in a weak field limit, and, 
consequently, far from the ABGB black hole as well as for  $Q/\mathcal{M} \ll 1$ the
line element resembles the Reissner-Nordstr\"om (RN) solution;
noticeable differences appear near the extremality limit. 
On the other hand, as $r \to 0$ one has the asymptotic behaviour
\begin{equation}
-g_{tt} =\frac{1}{g_{rr} }\sim 1-\frac{4\mathcal{M}}{r}\exp \left( \frac{-Q^{2}}
{\mathcal{M}r}\right)
\end{equation}
and this very information suffices to demonstrate the finiteness 
of the curvature invariants. 

Although
more complicated than the RN geometry the ABGB solution allows exact
analytical treatment. Indeed, the location of the horizons can be
expressed in terms of known special functions that certainly
simplifies investigations of its causal structure. The ABGB geometry
has been studied by a number of authors and from various points of
view. See, for example~\cite{BretonN,Radinschi,Yang,MatMPLA,Bur1,Bur2,Kocio1,Berej,
JM2004,JaSpin,Try}, where the stability of the ABGB spacetime, its 
gravitational energy, generalizations and
the stress-energy tensor of the quantized field propagating in such a
geometry have been analyzed. 
Especially interesting are the thermodynamic considerations
presented in Refs.~\cite{Kor,JaActa}.

In this paper we shall generalize the ABGB solution to the cosmological
background. Although the solution is valid for any $\Lambda$ we shall
restrict ourselves to the positive cosmological constant.
Consequently, an interesting group of related solutions describing
topological black holes are not considered here. The paper is
organized as follows: In section 2 we construct the solution
describing the ABGB black hole in the de Sitter geometry and give its
qualitative discussion. The quantitative discussion of its main
features is contained in Sec.3. The near-horizon geometry of the
degenerate configurations is constructed and discussed in Sec.4
whereas the analysis of the lukewarm black holes 
is presented in Sec.5. Finally, Sec.6 contains short
discussion and suggestions for extending this work. Throughout the
paper the geometric system of units is used and our conventions follow
the  conventions of MTW.

\section{Cosmological solutions of the coupled system of equations 
of nonlinear electrodynamics and gravity}

 In the presence of the cosmological constant the coupled system of
the nonlinear electrodynamics and gravity is described by the action 
\begin{equation}
S=\frac{1}{16\pi G}\int \left( R-2\Lambda \right) \sqrt{-g}\,d^{4}x+S_{m},
\end{equation}
where 
\begin{equation}
S_{m}=-\dfrac{1}{16\pi }\int \mathcal{L}\left( F\right) \sqrt{-g}\,d^{4}x.
\end{equation}
Here $\mathcal{L}\left( F\right) $ is some functional of $F=F_{ab}F^{ab}$
(its exact form will be given later) and all symbols have their usual
meaning. The tensor $F_{ab}$ and its dual $^{*}F^{ab}$  satisfy
\begin{equation}
\nabla _{a}\left( \dfrac{d\mathcal{L}\left( F\right) }{dF}F^{ab}\right) =0
\end{equation}
and
\begin{equation}
\nabla _{a}\,^{\ast }F^{ab}=0.
\end{equation}
 The stress-energy tensor defined as 
\begin{equation}
T^{ab}=\frac{2}{\sqrt{-g}}\frac{\delta }{\delta g_{ab}}S_{m}  \label{tensep}
\end{equation}
is given therefore by 
\begin{equation}
T_{a}^{b}=\dfrac{1}{4\pi }\left( \dfrac{d\mathcal{L}\left( F\right) }{dF}%
F_{ca}F^{cb}-\dfrac{1}{4}\delta _{a}^{b}\mathcal{L}\left( F\right) \right) .
\end{equation}
                                
Let us consider the spherically symmetric and static configuration described
by the line element of the form 
\begin{equation}
ds^{2}=-e^{2\psi \left( r\right) }f(r)dt^{2}+\frac{dr^{2}}{f(r)}%
+r^{2}d\Omega ^{2},  \label{el_gen}
\end{equation}
where $f(r)$ and $\psi(r)$ are unknown functions. 
The spherical symmetry places restrictions on the components of $F_{ab}$
tensor and its only nonvanishing components compatible with the assumed
symmetry are $F_{01}$ and $F_{23}$. Simple calculations yield 
\begin{equation}
F_{23}=Q\sin \theta
\end{equation}
and 
\begin{equation}
r^{2}e^{-2\psi }\dfrac{d\mathcal{L}\left( F\right) }{dF}F_{10}=Q_{e},
\end{equation}
where $Q$ and $Q_{e}$ are the integration constants interpreted as the
magnetic and electric charge, respectively. In the latter we shall assume
that the electric charge vanishes, and, consequently, $F$ is given by 
\begin{equation}
F=\dfrac{2Q^{2}}{r^{4}}.  \label{postacF}
\end{equation}
The stress-energy tensor (\ref{tensep}) calculated for this configuration
is 
\begin{equation}
T_{t}^{t}=T_{r}^{r}=-\dfrac{1}{16\pi }\mathcal{L}\left( F\right)  \label{t1}
\end{equation}
and 
\begin{equation}
T_{\theta }^{\theta }=T_{\phi }^{\phi }=\dfrac{1}{4\pi }\dfrac{d\mathcal{L}%
\left( F\right) }{dF}\dfrac{Q^{2}}{r^{4}}-\dfrac{1}{16\pi }\mathcal{L}\left(
F\right),
\end{equation}
which reduces to its Maxwell form for $\mathcal{L}(F) = F.$
With the substitution
\begin{equation}
f(r)=1-\frac{2M(r)}{r}
                     \label{fM}
\end{equation}
the left hand side of the time and radial components of the Einstein field equations 
\begin{equation}
L_{a}^{b}\equiv G_{a}^{b}+\Lambda \delta_{a}^{b} = 8 \pi T_{a}^{b}
\end{equation}
assume simple and transparent form
\begin{equation}
L_{t}^{t}=-\frac{2}{r^{2}}\frac{dM}{dr}+\Lambda ,\hspace{0.4cm}
L_{r}^{r}=L_{t}^{t}+\frac{2}{r}\left( 1-\frac{2M}{r}\right) \frac{d\psi }{dr}
+\Lambda, \label{1st}
\end{equation}
and the resulting equations can be easily (formally) integrated.

Further considerations require specification of the Lagrangian $\mathcal{L}
\left( F\right) .$ 
We demand that it should have proper asymptotic, i.e., in a weak field limit it should approach
$F.$
Following Ay\'on-Beato, Garc\'\i a~\cite{abg} and Bronnikov~\cite{kiryll1} 
let us
chose it in the form 
\begin{equation}
\mathcal{L}\left( F\right) \,=F\left[ 1-\tanh ^{2}\left( s\,\sqrt[4]{\frac{
Q^{2}F}{2}}\right) \right] ,  \label{labg}
\end{equation}
where 
\begin{equation}
s=\frac{\left| Q\right| }{2b},  \label{sabg}
\end{equation}
and the free parameter $b$ will be adjusted to guarantee regularity at the
center.
Inserting Eq.~(\ref{sabg}) into (\ref{labg}) and makig use of Eq.~(\ref
{postacF}) one has 
\begin{equation}
8\pi T_{t}^{t}=8\pi T_{r}^{r}=-\frac{Q^{2}}{r^{4}}\left( 1-\tanh ^{2}\frac{
Q^{2}}{2br}\right) .  \label{tep}
\end{equation}
Now the equations can easily be integrated in terms of the
elementary functions:
\begin{equation}
M\left( r\right) =C_{1}-b\tanh \frac{Q^{2}}{2br}+\frac{\Lambda r^{3}}{6},
\hspace{0.4cm}\psi \left( r\right) =C_{2}  \label{mm0}
\end{equation}
where $C_{1}$ and $C_{2}$ are the integration constant. Making use of the
conditions 
\begin{equation}
M(\infty)={\mathcal M},   \hspace{0.4cm}     \psi(\infty)=0
\end{equation}
gives $C_{1}=\mathcal{M}$ and $C_{2} = 0.$ On the other hand, demanding
the regularity of the line element as $r\rightarrow 0$ yields $b_{1}=%
\mathcal{M,}$ and, consequently, the resulting line element has 
the form (\ref{el_gen}) with $\psi(r)=0$ and
\begin{equation}
f(r) = 1-\frac{2 \mathcal{M}}{r}\left(1-\tanh\frac{Q^{2}}{2 \mathcal{M} r}
\right)-\frac{\Lambda r^{2}}{3}.
        \label{el_gen1}
\end{equation}
We shall call this solution the Ay\'on-Beato-Garc\'\i a-Bronnikov-de Sitter 
solution (ABGB-dS).
It could be easily shown that putting $Q=0$ yields the Schwarzschild-de Sitter
(Kottler) solution, whereas for $\Lambda=0$ one gets the Ay\'on-Beato, Garc\'\i a  
line element as reinterpreted by Bronnikov (ABGB).

To study ABGB-dS line element it is convenient to introduce the
dimensionless quantities $x=r/M$, $q=\left| Q\right| /M$ and $\lambda
=\Lambda M^{-2}$. For $\lambda > 0$ the equation 
\begin{equation}
1-\frac{2}{x}\left( 1-\tanh\frac{q^2}{2x}\right) -\frac{1}{3}\lambda x^2 =0
                                       \label{eqq}
\end{equation}
has, in general, four
roots; the case $\lambda=0$ has been treated analytically 
in Refs~\cite{Kocio1,Berej,JM2004}.
Unfortunately, Eq.~(\ref{eqq}) cannot be solved in terms of known
transcendental functions and one is forced to refer to some
approximations or employ numerical methods. Simple analysis indicate that
for $x>0$ it can have, depending on the
values of the parameters, three, two or one distinct real solutions.
The above configurations can, therefore,
have three distinct horizons located at zeros of $f\left( r\right) $, a
degenerate and a nondegenerate horizon, and, finally, one triply
degenerate horizon. 
Let us consider each of the configuration more closely
and denote the inner, the event and the cosmological horizon by $r_{-}$, $
r_{+}$ and $r_{c},$ respectively. The first configuration is characterized
by $r_{-}<r_{+}<r_{c}$. The second configuration can be realized in
two different ways depending on which horizons do merge and is characterized
either by \ $r_{-}=r_{+}<r_{c}$ (degenerate horizons of the first type, referred to as
the cold black hole) 
or $r_{-}<r_{+}=r_{c}$ (degenerate horizons of the 
second type sometimes referred to
as the charged Nariai black hole~\footnote{It must not be confused with 
the charged Nariai solution which will be discussed in section 4.}). 
Finally, for the third (ultracold)
configuration one has $r_{-}=r_{+}=r_{c}.$ The degenerate horizons are
located at simultaneous zeros of $f\left( r\right) $ and $f^{\prime }\left(
r\right) $ for the cold or the Nariai black hole and $f\left( r\right) =f^{\prime }\left( r\right)= 
f^{\prime \prime }\left( r\right)=0 $ for the ultracold black hole. 
Additionally one can single out the lukewarm configuration, for which the
Hawking temperature of the black hole equals the temperature of the
cosmological horizon.

The Penrose diagrams visualizing  two-dimensional sections of the
conformally transformed ABGB-dS geometry is precisely of the type considered earlier
for the Reissner-Nordstr\"om-de Sitter black hole~\cite{Mellor} with the one notable distinction:
the central singularity must be replaced be a regular region.

Although the line element (\ref{el_gen}) with (\ref{el_gen1}) 
is rather complicated and cannot be studied
analytically one can easily analyze its main features simply by referring to
its important limits. First, let us observe that for small $q$ $\left(
q\ll 1\right) $ as well as at great distances form the center $\left(
x\gg 1\right) $ the ABGB-dS solution closely resembles that of RN-dS. Indeed,
expanding $f\left( r\right) $ one obtains 
\begin{equation}
f\,=1-\frac{2\mathcal{M}}{r}+\frac{Q^{2}}{r^{2}}\,-\frac{\Lambda r^{2}}{3}-\,%
\frac{Q^{6}}{12\mathcal{M}^{2}r^{4}}\,+\,....
\end{equation}
On the other hand, as $r\rightarrow 0$, one has 
\begin{equation}
f\sim 1-\frac{4\mathcal{M}}{r}\exp \left( \frac{-Q^{2}}{\mathcal{M}r}\right)
-\frac{\Lambda r^{2}}{3}
\end{equation}
and the metric in the vicinity of the center may by approximated by the
de Sitter line element. One concludes, therefore, that the solution is regular 
at $r=0$ and, in a view of the asymptotic behaviour of the line element,
to demonstrate this it is unnecessary to calculate the curvature 
invariants explicitly. For example, at $r=0$ the Kretschmann scalar is equal $8\Lambda^{2}/3,$
as expected.  Further, observe that for small $\lambda $ the
structure of the ABGB-dS geometry is to certain extend qualitatively similar
to the ABGB spacetime. Simple analysis indicates that there are, at most,
three positive roots of the equation $f\left( r\right) =0.$ Two of them  
are located closely to the
inner and event horizons of the ABGB black hole whereas the third one,
located approximately at 
\begin{equation}
x_{c}\simeq \left( \frac{3}{\lambda }\right) ^{1/2}
                              \label{cosmo}
\end{equation}
is to be interpreted as the cosmological horizon.

\section{Horizon structure of the ABGB-dS black holes}
Having established the
main features of the ABGB-dS solution qualitatively let us study it in
more detail and consider the approximate solutions for $\lambda \ll 1$ first.
We shall start, however, with a brief discussion of the $\lambda =0$ case
and present the results that will be needed in the subsequent calculations.
In Ref.~\cite{Kocio1} it has been shown that for $\lambda =0$ 
the location of the inner, 
$r_{-}^{(0)},$ and event horizon, $r _{+}^{(0)},$ of the ABGB 
black holes can be
expressed in terms the real branches of the Lambert functions, $W_{\left(
\pm \right) }\left( s\right) $: 
\begin{equation}
\rho _{\pm }=r^{(0)}_{\pm}/{\mathcal M}=-\frac{4q^{2}}{4W_{\left( \pm \right) }\left( -\frac{q^{2}}{4}%
\exp \left( q^{2}/4\right) \right) -q^{2}}.
\end{equation}
Here $W_{+}$ (the principal branch) and $W_{-}$ are the only real branches 
of the Lambert function.
Simple manipulations shows that $\rho _{+}$ and $\rho _{-}$ for 
\begin{equation}
q=q_{c}=2\sqrt{W_{+}\left( 1/e\right) }\equiv 2\sqrt{w_{0}}
                      \label{abg_q}
\end{equation}
merge at 
\begin{equation}
\rho _{c}=\frac{4w_{0}}{1+w_{0}}.
                      \label{abg_x}
\end{equation}
For $q^{2}>q_{c}^{2}$ the degenerate solution bifurcate into a pair of two
complex roots. 

For small $\lambda$ one expects that the inner and the event horizon
lies closely to the $\rho_{-}$ and $\rho_{+},$ respectively, and the position 
of the cosmological horizon can always be approximated by Eq. (\ref{cosmo}). 
Depending on $q$
there will be two horizons located at the roots $x_{-}$ and $x_{+}$, which
for $q^{2}=q_{cr}^{2}$ coalesce into the degenerate horizon $x_{cr}.$ For $%
q^{2}>q_{cr}^{2}$ there are no real solutions for $x_{\pm}$ and $x_{c}$ tends
to $(3/\lambda)^{1/2}$ with increasing $q.$

Now, let us consider the cold black hole, i.e. the one for which $
x_{-}=x_{+}=x_{cr}.$ Taking $\lambda $ to be a small parameter of the
expansion one obtains
\begin{eqnarray}
q_{cr}^{2} &=&4w_{0}+\frac{64w_{0}^{3}}{3\left( 1+w_{0}\right) ^{2}}\lambda +%
\frac{1024w_{0}^{5}\left( 5+3w_{0}\right) }{9\left( 1+w_{0}\right) ^{5}}%
\lambda ^{2} \nonumber\\
&&+\frac{32768}{81\left( 1+w_{0}\right) ^{8}}\left(
59+65w_{0}+18w_{0}^{2}\right) \lambda ^{3}+O\left( \lambda ^{4}\right)
\end{eqnarray}
and

\begin{eqnarray}
x_{cr} &=&\frac{4w_{0}}{1+w_{0}}+\frac{64w_{0}^{3}\left( 3+w_{0}\right) }{%
3\left( 1+w_{0}\right) ^{4}}\lambda +\frac{1024w_{0}^{5}\left(
25+18w_{0}+3w_{0}^{2}\right) }{9\left( 1+w_{0}\right) ^{7}}\lambda ^{2}\nonumber \\
&&+\frac{32768w_{0}^{7}}{81\left( 1+w_{0}\right) 10}\left(
413+461w_{0}+162w_{0}^{2}+18w_{0}^{3}\right) \lambda ^{3}+O\left( \lambda
^{4}\right) .
                             \label{bb}
\end{eqnarray}
For $q^{2}<q_{cr}^{2},$ following Romans~\cite{Romans}, we shall introduce the
dimensionless parameter $\Delta =\sqrt{q_{cr}^{2}-q^{2}\text{ }}$ and look
for solutions of Eq. (\ref{eqq}) of the form
\begin{equation}
x_{\pm }=\rho_{\pm} +x_{1}^{\left( \pm \right) }\lambda
+x_{2}^{\left( \pm \right) }\lambda ^{2}+O\left( \lambda ^{3}\right) .
                            \label{aa}
\end{equation}
where 
\begin{equation}
\rho_{\pm}=\frac{4\left( q_{c}^{2}-\Delta ^{2}\right) }{
4W_{\pm }\left( \eta \right) -q_{c}^{2}+\Delta ^{2}}
\end{equation}
and
\begin{equation}
\eta = \frac{\Delta^{2}-q_{c}^{2}}{4}\exp\left(\frac{q^{2}_{c}-\Delta^{2}}{4}\right).
\end{equation}
Now, solving a chain of equations of ascending complexity, one has
\begin{equation}
x_{1}^{\left( \pm \right) }=\frac{4\rho_{\pm}\left[ 64w_{0}^{3}-16w_{0}^{3}\rho_{\pm}-\left( 1+w_{0}\right)
^{2}\rho_{\pm}^{3}\right] }{3\left( 1+w_{0}\right) ^{2}\left[ \left(
4-\rho_{\pm}\right) \left( 4w_{0}-\Delta ^{2}\right) -4\rho_{\pm}\right] }
\end{equation}
and
\begin{eqnarray}
x_{2}^{\left( \pm \right) } &=&\frac{4}{\left( 4-\rho_{\pm}\right) \left[ \left( 4-\rho_{\pm}\right) \left(
4w_{0}-\Delta ^{2}\right) -4\rho_{\pm}\right] }\left[ 2 \left(x_{1}^{\left( \pm \right) }\right)^{2}+\frac{1}{9}
\left( \rho_{\pm}-2\right) \rho_{\pm}^{6}\right. \nonumber \\
&&\left. +\frac{2}{3}\left( \rho_{\pm}-6\right) \rho_{\pm}^{3}x_{1}^{\left( \pm \right) }+\frac{
256\rho_{\pm}w_{0}^{5}}{9\left( 1+w_{0}\right) ^{5}}\left( 5+3w_{0}\right) \left(
\rho_{\pm}-4\right) ^{2}\right].
\end{eqnarray}
It could be easily shown that putting $\Delta =0$ and taking limit of (\ref{aa}) as $
\rho_{\pm}\rightarrow 4w_{0}/\left( 1+w_{0}\right) $ one
obtains (\ref{bb}).

In the situations when the cosmological constant cannot be regarded as
small, the analytical treatment of the horizon structure of the ABGB-dS black holes is 
impossible. However, although we are unable
to calculate the exact location of horizons in the spacetime of ABGB-dS black holes,
one can use a simple trick. Indeed, the form of the equation~(\ref{eqq})
suggests that it can be
solved easily with respect to $q$ yielding
\begin{equation}
q =\pm \sqrt{x \ln \frac{12-3x+ \lambda x^{2}}{x(3 - \lambda x^{2})}}.
                                 \label{inv}
\end{equation}
This allows to draw curves $q=q(x)$ for various (constant) $\lambda .$
The extrema of the curves represent either the degenerate horizons of the cold black holes
or the charged Nariai black holes. A closer examination shows that 
for $q>0$ one has minima for the configurations with $r_{+}=r_{c}$ and maxima
for the cold black hole. Drawing, on the other hand, $\lambda = \lambda(x)$ curves
for constant values of $q$ one has minima for the configurations with $r_{-} =r_{+}$
and maxima for the charged Nariai black holes.
Except the ultracold black hole the
configurations with the cosmological horizon only are not considered in this paper.
The results of such a calculation is displayed in Fig.1.
\begin{figure}[h]
\centering
\includegraphics[width=8cm]{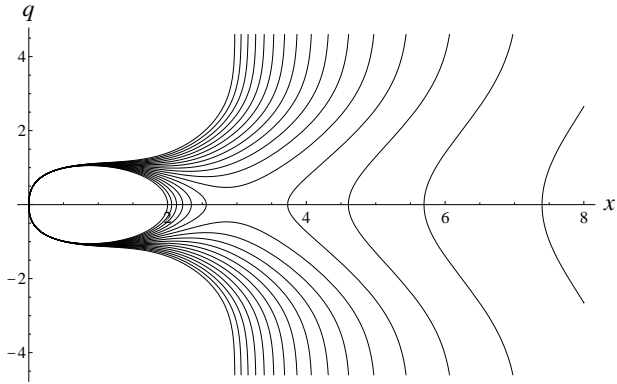}
\caption{The curves in this figure display values of $q$ as function of $x$
(where $x$ denotes the positive roots of Eq.(\ref{eqq})) for constant $\lambda$
(see Eq.(\ref{inv})).
The curves are drawn for $\lambda=0.02 i$ for $i=0,...,15$ The extrema 
of curves represent degenerate configurations.}
\end{figure}
Now, rotating the diagram counter-clockwise by ninety degrees
and subsequently reflecting the thus obtained curves by 
the vertical axis one gets the desired result.
On the other hand one can employ numerical methods and the results 
of such calculations are presented in Fig. 2,
where, for better clarity, only $q>0$ region has been displayed. 
To investigate the horizon structure it is useful to focus attention on
$x=x(q)$ curves of constant  $\lambda,$ where $x$ is, depending on 
its position on the curve, one the three horizons.
For each $\lambda \geq \lambda_{0}$ (where $\lambda_{0}$ denotes some critical value 
of the cosmological constant to be given later) the (rescaled) radii of the inner
horizon (the lower branch), the event horizon (the middle branch) and the 
cosmological horizon (the upper branch) comprise an S-shaped curve
and the turning points of each curve represent degenerate horizons.
On the other hand, for $\lambda <\lambda_{0}$ there is only one turning point representing
cold black hole with $r_{+}=r_{-}$ and the cosmological horizon branch is separated form the rest 
of the curve. The degenerate horizon of the second
type appears precisely for $\lambda_{0} = 1/9.$ 
\begin{figure}
\centering
\includegraphics[width=8cm]{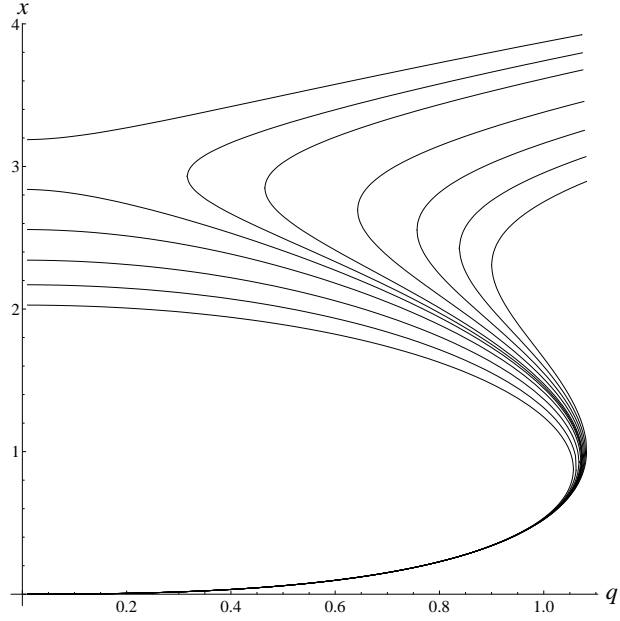}
\caption{The positive roots of Eq. (\ref{eqq}) as a function of $q.$
Bottom to top the curves are drawn for $\lambda =0.008, 0.02, 0.05, 0.08,
0.1, 0.11, 0.12, 013, 0.14, 0.15, 016.$
The lower branches represent the inner horizon, which, for $q<1$
is practically insensitive to the changes of the cosmological constant.
For $\lambda >1/9$ there are two additional branches representing the event
and the cosmological horizon comprising S-shaped curves. For $\lambda <1/9$
the upper branch (the cosmological horizon) is disjoint from the rest of the curve
(consisting of the lower and middle branches) ,and,
for small $\lambda,$ it is located approximately at $\sqrt{3/\lambda}.$}
\end{figure}

Although the qualitative behaviour of the degenerate horizons as function of the
cosmological constant, such as rather weak dependence of the location of the 
degenerate horizon of the first type, may easily be inferred from the above 
analysis, we shall discuss it in more detail. 
The dependence of the location of the degenerate horizons as functions
of $\lambda$ can be calculated from Eq.~(\ref{inv}) and the results
are displayed in Fig 3. The lower branch represents
degenerate horizons of the first type whereas the upper one represents 
degenerate horizons of the second kind, and, finally, the branch point represents the 
triply degenerated horizon. Such a configuration occurs at
$x=1.34657$ for $\lambda =0.246019$ and $|q|= 1.1082.$

\begin{figure}
\centering
\includegraphics[width=8cm]{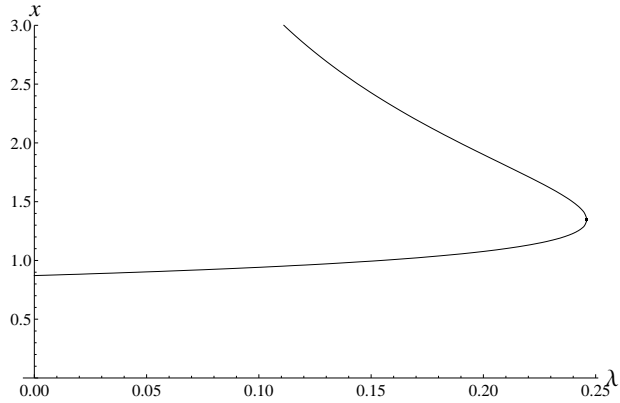}

\caption{The radii of the extremal horizons as function of $\lambda.$ The lower branch
represents the extreme horizons of the cold black hole $(x_{-} =x_{+})$ whereas the upper branch
represents the extreme horizons of the second type $(x_{+}=x_{c}).$ 
The branch point represents ultracold black hole $(x_{-} =x_{+}=x_{c})$ and the point (1/9, 3) 
of the upper branch represents the extreme Schwarzschild-de Sitter solution.}
\end{figure}
\section{Extreme configurations}

The ABGB-dS solution gives rise to a number of solutions constructed
by applying some limiting procedure in the near extreme geometry. For
example, it is a well known fact that the near horizon geometry of the
Reissner-Nordstrom solution is properly described by the Bertotti-
Robinson line element~\cite{Brandon,Bert,Rob,Paul1} whereas Ginsparg
and Perry~\cite{Ginsparg} demonstrated that the extreme Schwarzschild-
de Sitter black hole is naturally connected with the Nariai
solution~\cite{Nariai1,Nariai2}. The procedure of Ginsparg and Perry
has been subsequently generalized and employed in a number of
physically interesting cases, such as C-metrics~\cite{Lemos0}, D-
dimensional black holes~\cite{Cald,Lemos1} and in construction of
various instantons~\cite{MannRo,Hawk}, to name a few. In this section we shall 
construct the exact solutions to
the Einstein field equations which are asymptotically congruent to the
near horizon geometry of the ABGB-dS black holes. 

First let us
consider the situation when the inner horizon is close to the event
horizon. For $r_{-} \leq r \leq r_{+}$ the function $f$ can be approximated 
by a parabola $\alpha (x-x_{+})(x-x_{-})$ and the degenerate horizon, $x_{d},$
by $(x_{+}+x_{-})/2.$ Putting $q^{2} =q_{d}^{2}-\varepsilon^{2} \Delta^{2},$
where $\varepsilon$ is a small parameter that measures deviation from 
the extremal configuration and $\Delta$ should be chosen is such a way
to guarantee $x_{\pm} =x_{d}\pm \varepsilon$ one can easily determine $\alpha.$
Indeed, it can be shown that for a given $\lambda$ one has
\begin{equation}
\alpha = 4 \frac{\partial f}{\partial q^{2}_{|q_{d}}}
\frac{\Delta^{2}\varepsilon^{2}}{(r_{+}-r_{-})^{2}}
= \frac{\partial f}{\partial q^{2}_{|q_{d}}}\Delta^{2} > 0
\end{equation}
Similarly, for $r_{+}\leq r \leq r_{c},$ one can approximate the function $f$ by 
a parabola $\beta (x-x_{+})(x-x_{c})$ and the degenerate horizon by $(x_{+}+x_{c})/2.$
Putting $q^{2} = q^{2}_{d} +\varepsilon^{2} \tilde{\Delta}^{2}$ one obtains 
\begin{equation}
\beta = - \frac{\partial f}{\partial q^{2}_{|q_{d}}}\tilde{\Delta}^{2} <0.
\end{equation}

We shall illustrate the procedure by the example of the ABGB black hole.
First, observe that making use of the expansions of the Lambert functions $W_{+}$ and $W_{-}$
\cite{Lambert1,Lambert2}
\begin{equation}
W_{\pm}(z) =-1 +p -\frac{1}{3}p^{2} +...,
\end{equation}
where $p=\sqrt{2 (ez+1)}$ for the principal branch and $p=-\sqrt{2 (ez+1)}$ for $W_{-}(z),$
one has
\begin{equation}
x_{\pm} =\frac{4w_{0}}{1+w_{0}} \pm \frac{\sqrt{8 w_{0}}}{(1+w_{0})^{3/2}}\Delta\varepsilon +...,
\end{equation} 
and, consequently, 
\begin{equation}
\Delta =\frac{(1+w_{0})^{3/2}}{\sqrt{8 w_{0}}}. 
     \label{ddd}
\end{equation}
(Notation has been slightly modified as compared to Sec. 3).
Further observe that 
\begin{equation}
\frac{\partial f}{\partial q^{2}_{|q_{d}}} = \frac{1}{4w_{0}}
    \label{ppp}
\end{equation}
and $f$ may be approximated by 
\begin{equation}
f=\frac{(w_{0}+1)^{3}}{32 w_{0}^{2}}(x-x_{-})(x-x_{+})=A (r-r_{-})(r-r_{+}).
\end{equation}
Finally, introducing new coordinates $\tilde{t} = T/(\varepsilon A),$
 $r = r_{d}+\varepsilon \cosh y,$ and taking  limit $\varepsilon \to 0$
 one obtains the line element in the form~\cite{JM2004}
\begin{equation}
ds^{2} = \frac{1}{A}\left(- \sinh^{2}y\, dT^{2} + dy^{2} \right)
+r_{d}^{2}\left(d\theta^{2} + \sin^{2}\theta \,d\phi^{2}\right).
        \label{ads2xs2}
\end{equation}
 Since the modulus of curvature 
radii of the maximally symmetric subspaces are different
this solution does not belong to the Bertotti-Robinson~\cite{Bert,Rob} class.
Topologically it is a product of the round two sphere of a constant radius and 
the two dimensional anti-de Sitter geometry. We will call this solution 
a generalized Bertotti-Robinson solution.

 Now, let us return to the ABGB-dS geometry 
and observe that $\frac{\partial f}{\partial q^{2}_{|q=q_{d}}}$
is always nonnegative. Since there are no analytical expressions describing 
the exact location of the horizons we shall employ the perturbative approach.
Starting with the configurations with $r_{-}$ close to $r_{+},$ 
and repeating the steps above, one obtains the line element (\ref{ads2xs2}) with 
\begin{equation}
A = \frac{\partial f}{\partial q^{2}_{|q_{d}}}\frac{\Delta^{2}}{M^{2}},
    \label{AA}
\end{equation}
where 
\begin{equation}
\Delta^{2} = 2\frac{q_{d}^{2}}{x_{d}^{2}} - \cosh^{2}\frac{q_{d}^{2}}{2 x_{d}} 
- \frac{q_{d}^{4}}{2 x_{d}^{3}}
\tanh\frac{q_{d}^{2}}{2 x_{d}}
    \label{DeltSq}
\end{equation}
and
\begin{equation}
 \frac{\partial f}{\partial q^{2}_{|q_{d}}}=\frac{1}{x_{d}^{2}
 \cosh^{2}\frac{q_{d}^{2}}{2 x_{d}}}.
    \label{poch}
\end{equation}
 The curvature scalar 
of the geometry (\ref{ads2xs2}) is a sum of curvatures of the maximally
symmetric subspaces
\begin{equation}
R = R_{AdS_{2}} + R_{S^{2}},
\end{equation}
where
$R_{AdS_{2}}=-2A$ and $R_{S^{2}}=2/r_{d}^{2}.$
We shall call (\ref{ads2xs2}) a generalized cosmological Bertotti-Robinson solution.
It can easily be checked that putting $q_{d}=q_{c}$ and $x_{d}=\rho_{c}$ 
as given by Eqs. (\ref{abg_q}) and (\ref{abg_x}), respectively, one obtains (\ref{ddd}) and (\ref{ppp}).

On the other hand, for the near extreme configurations with $r_{+}$ close to $r_{c}$ 
we shall put $q^{2} =q_{d}^{2}+\varepsilon^{2} \tilde{\Delta}^{2}.$ 
Repeating calculations one has $\tilde{\Delta}^{2} =-\Delta^{2}.$ It should be noted however
that for $x_{d}=x_{+}=x_{c}$ the parameter $\tilde{\Delta}^{2}$  is positive and hence
$\beta$ is negative, as required. Introducing new coordinates
$\tilde{t} = T/(\varepsilon B)$
and $r = r_{d}+\varepsilon \cos y,$ in the limit $\varepsilon \to 0,$ one obtains
\begin{equation}
ds^{2} = \frac{1}{B}\left(- \sin^{2}y\, dT^{2} + dy^{2} \right)
+r_{d}^{2}\left(d\theta^{2} + \sin^{2}\theta \,d\phi^{2}\right),
                \label{ds2xs2}
\end{equation}
where $ B =-\beta.$ 
Topologically it is a product of the round two sphere and the two dimensional de Sitter  spacetime
and the curvature scalar is given by
\begin{equation}
R = R_{dS_{2}} + R_{S^{2}},
\end{equation}
where $ R_{dS_{2}}=2 B.$
We shall call this solution a generalized Nariai solution.

Differentiating the function $f$ with respect to $x$ twice,
subtracting $2 f(x)/x^2 =0$ and dividing thus obtained result
by 2 one concludes that
\begin{equation}
A =\frac{1}{2} f''(r_{d}).
\end{equation}
It follows then that $A$ vanishes at the ultraextremal horizon.
Finally observe that putting $y =\xi A^{1/2}$ in (\ref{ads2xs2}) and taking the limit
$A \to 0,$ one obtains
\begin{equation}
ds^{2} = -\xi^{2} dT^{2} +d\xi^{2} + r_{d}^{2} \left(d\theta^{2} + \sin^{2}\theta \,d\phi^{2}\right).
\end{equation}
Topologically it is a product of the two-dimensional Minkowski space and the round 
two-sphere of fixed radius
and can be identified with the Pleba\'nski-Hacyan~\cite{Plebanski,Ortaggio,OrtaggioP}
solution.

Although we have adopted the point of view that the cosmological constant 
is not a parameter in the solutions space but, rather,  the parameter 
of the space of theories, one can equally well keep $q$ constant and 
change $\lambda.$ Indeed, repeating the calculations with 
$\lambda = \lambda_{d}+\varepsilon^{2}\Delta^{2}$ for $r_{-}$ 
close to $r_{+}$ and $\lambda = \lambda_{d}-\varepsilon^{2}\Delta^{2}$
for $r_{+}$ close to $r_{c}$ one obtains precisely (\ref{ads2xs2}) and (\ref{ds2xs2}), 
respectively. The sign can be deduced form the analysis of the $\lambda=\lambda(x)$ curves
obtained from Eq. (\ref{eqq}), and,
as before, the subscript $d$ denotes degenerate configurations.
 
Since the $AdS_{2}\times S^{2}$ and $dS_{2}\times S^{2}$  (with arbitrary radii of 
the maximally symmetric subspaces) appear to describe universally the geometry of 
the vicinity of the (doubly) degenerate horizons, on can use this information in construction 
of the coefficients $A$ and $B.$ Indeed, observe that $f''(r_{d})>0$ at the degenerate
horizon of the cold black hole, whereas $f''(r_{d})<0$  at the degenerate horizon 
of the second type. At the degenerate horizons the Einstein field equations reduce to 
\begin{equation}
-\frac{1}{r_{d}^{2}}+ \Lambda = 8\pi T_{t}^{t}
\end{equation}
and
\begin{equation}
\frac{1}{2}f''(r_{d}) + \Lambda = 8\pi T_{\theta}^{\theta}
\end{equation}
Consequently, $A =f''(r_{d})/2$ and $B=-f''(r_{d})/2$
and at the degenerate horizon of the ultracold configuration
one has $f''(r_{d}) =0.$

\section{Lukewarm configuration}
Finally, let us consider the important case of the lukewarm black holes,
i.e. the black holes with the Hawking temperature of the event horizon
equal to that associated with the  cosmological horizons.
From the point of view of the quantum field theory in curved background the lukewarm
black holes are special. It has been shown that for the two-dimensional models 
it is possible to construct a regular thermal state~\cite{Lizzie1}. Moreover, recent
calculations of the vacuum polarization indicate that it is regular on both
the event and cosmological horizons of the D=4 lukewarm RN-dS black holes~\cite{Lizzie2}.

\begin{figure}
\centering
\includegraphics[width=8cm]{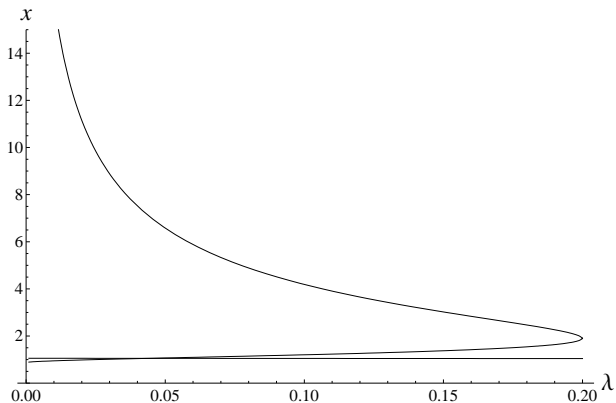}
\caption{The radii of horizons of the lukewarm black hole as functions of $\lambda.$ 
The branch point should be excluded as it represents the degenerate configuration 
of the second type.  The almost horizontal line displays the values of $q.$
}
\end{figure}

As the lukewarm black holes are characterized by the condition $T_{H} = T_{C}$ 
the radii of the horizons with this property satisfy the system of equations
\begin{equation}
f(r_{+})=f(r_{c})=0,\hspace{5mm}   f'(r_{+})+f'(r_{c})=0.
\end{equation}
Since one expects that the structure of the horizons of the ABGB-dS black hole is qualitatively
similar to its maxwellian counterpart it is worthwhile to analyze briefly
lukewarm Reissner-Nordstr\"om black holes. Simple calculations indicate that such configurations
are possible for $q=1$ and the locations of the event and the cosmological horizons are 
given by 
\begin{equation}
x_{+} =\frac{{\it l}}{2}\left(1-\sqrt{1-\frac{4M}{{\it l}} }   \right) 
\end{equation}
and
\begin{equation}
x_{c} =\frac{{\it l}}{2}\left(1+\sqrt{1-\frac{4M}{{\it l}} }   \right), 
\end{equation}
where ${\it l}=\sqrt{3/\lambda}.$ For $\lambda =3/16$ $r_{+}$ and $r_{c}$
coalesce into the degenerate horizon of the second type.
One expects, therefore, that for the lukewarm ABGB-dS black holes $q$ should always
be close to 1. Results of our numerical calculations are  presented in Fig.4., where
the (rescaled) radii of the event horizon (the lower branch) and the cosmological
horizon (the upper branch) are drawn. The function $q=q(\lambda)$ is  almost horizontal
The branch point should be excluded as it refers to the ultracold black holes.

\section{Final Remarks}
In this paper we have constructed the regular solution to the system of coupled equations 
of the nonlinear electrodynamics and gravity in the presence of the cosmological 
constant. We have restricted to the positive $\Lambda$ and concentrated on the
classical issues, such as location of the horizons, degenerate configurations
and various solutions constructed form the two dimensional maximally symmetric 
subspaces. The discussion of the solutions with $\Lambda <0,$ both spherically symmetric
and topological,  have been intentionally omitted.
Outside the event horizon the ABGB-dS solution closely resembles RNdS; important differences appear,
as usual, for the near extreme configurations.
At large distances as well as in the closest vicinity of the center
the line element may be approximated by the de Sitter solution.

We indicate a few possible directions of investigations. First, it would 
be interesting to examine the vacuum polarization effects in the ABGB-dS
geometries and compare them with the analogous results calculated in
the RNdS spacetime. It should be noted in this regard that the geometries 
naturally connected with the ABGB-dS geometry, namely the generalized Bertotti-Robinson and
the cosmological charged Nariai metric are the exact solutions of the semiclassical 
Einstein field equations~\cite{Sahni1,Sahni2,Sahni3,Ollo,Solodukhin,Ja2000,JaO1}. 
Moreover, the interior of the ultraextremal ABGB-dS
black hole provides a natural background for studies initiated in Ref.~\cite{JaO2}.
This group of problems is actively investigated and the results will be published
elsewhere.


\end{document}